\input{epsf}
\documentstyle[times]{mn}

\title[Dynamical fragmentation]{Gas-cooling by dust during 
dynamical fragmentation}

\author[A. P. Whitworth, H. M. J. Boffin and N. Francis]
{A. P. Whitworth\thanks{e-mail: a.whitworth@astro.cf.ac.uk},
H. M. J. Boffin\thanks{e-mail: h.boffin@astro.cf.ac.uk}
and N. Francis\thanks{e-mail: n.francis@astro.cf.ac.uk}\\
Department of Physics \& Astronomy, University of Wales, 
Cardiff, CF2 3YB
}

\def\plotone#1{\centering \leavevmode \epsfxsize=\columnwidth \epsfbox{#1}}

\newcommand{\goo}{\stackrel{>}{\sim}}
\newcommand{\loo}{\stackrel{<}{\sim}}
\newcommand{\du}{_{\mbox{\tiny dust}}}
\newcommand{\mi}{_{\mbox{\tiny min}}}
\newcommand{\ma}{_{\mbox{\tiny max}}}
\newcommand{\ci}{_{\mbox{\tiny crit}}}
\newcommand{\co}{_{\mbox{\tiny comp}}}

\newcommand{\gd}{_{\mbox{\tiny g-d}}}
\newcommand{\ra}{_{\mbox{\tiny rad}}}
\newcommand{\CR}{_{\mbox{\tiny CR}}}
\newcommand{\Je}{_{\mbox{\tiny J}}}
\newcommand{\bg}{_{\mbox{\tiny o}}}

\newcommand{\X}{_{\mbox{\tiny X}}}
\newcommand{\Jh}{J_{\mbox{\tiny highest}}}
\newcommand{\CO}{_{\mbox{\tiny CO}}}

\begin{document}

\maketitle

\begin{abstract}
We suggest that the abrupt switch, 
from hierarchical clustering on scales $\goo$ 0.04 pc, 
to binary (and occasionally higher multiple) 
systems on smaller scales, which Larson 
has deduced from his analysis of the grouping of 
pre--Main-Sequence 
stars in Taurus, arises because pre-protostellar gas becomes 
thermally coupled to dust at sufficiently high densities. 
The resulting change from gas-cooling by molecular lines 
at low densities to gas-cooling by dust at high 
densities enables the matter to radiate much more 
efficiently, and hence to undergo dynamical fragmentation.

We derive the domain where gas-cooling by dust facilitates dynamical fragmentation. Low-mass ($\sim M_\odot$) clumps -- those supported mainly by thermal pressure -- can probably access this domain spontaneously, albeit rather quasistatically, provided they exist in a region where external perturbations are few and far between. More massive clumps probably require an impulsive external perturbation, for instance a supersonic collision with another clump, in order for the gas to reach sufficiently high density to couple thermally to the dust. Impulsive external perturbations should promote fragmentation, by generating highly non-linear substructures which can then be amplified by gravity during the subsequent collapse.
\end{abstract}


\section{Introduction}

Dynamical fragmentation (i.e. the growth of substructure 
during near-freefall collapse) is widely believed to play 
an important role in star formation, both as a means of 
forming star clusters, and as a means of forming binary 
systems (e.g. Bate, Bonnell \& Price 1995; Bodenheimer, 
1978; Bodenheimer \& Boss, 1981; Bonnell \& Bastian 1993; 
Bonnell \& Bate 1994; Boss, 1991, 1993; Boss \& 
Bodenheimer 1979; Burkert \& Bodenheimer 1996; Chapman et al.,  
1992; Hoyle, 1953; Larson, 1972; Miyama, 1992; Miyama, 
Hayashi \& Narita, 1984; Monaghan, 1994; Monaghan \& 
Lattanzio, 1991; Myhill \& Kaula, 1992; Nelson \& Papaloizou, 
1993; Pringle, 1989; Sigalotti \& Klapp, 1994; Turner et al., 1995; 
Whitworth et al., 1995).

Recently Larson (1995) has analyzed the grouping of 
pre--Main-Sequence (pre-MS) stars 
in Taurus, and identified a critical length-scale 
$L\ci \, \sim$ 
0.04 pc. On larger scales the pre-MS stars appear to be clustered 
in a self-similar hierarchy, with the mean surface-density 
of neighbours $N$ decreasing with angular separation $\theta$ as 
$N \, \propto \, \theta^{-0.62}$. On smaller scales the pre-MS stars have 
on average just one companion, which is usually tightly 
bound in a binary system; here the mean surface density of companions 
decreases as $N \, \propto \, \theta^{-2.15}$. 
Larson suggests that these binary 
(and occasionally higher multiple) systems are formed by 
dynamical fragmentation. He proposes that $L\ci$ is 
simply the Jeans length appropriate for the conditions 
prevailing in Taurus, and 
that the corresponding Jeans mass is 
$\sim M_\odot$. He then points out that if there is a 
preponderance of marginally 
Jeans-unstable clumps with $M \sim M_\odot$ collapsing 
and fragmenting into binary and triple systems, this will yield 
protostellar masses in the range 0.3 to 0.5 $M_\odot$, 
i.e. very close to the characteristic 
mass of the Initial Mass Function, $\sim 0.4 M_\odot$. 

However, 
there is a large range of densities in Taurus, 
and presumably an even large range of conditions in the 
various star-formation regions which form the entire 
ensemble of stars defining the Initial Mass Function. 
Therefore there must be a wide range of Jeans 
lengths.

Here we argue that the onset 
of fragmentation at the scale $L\ci$ is in fact due to the  
density becoming just high enough for the gas to couple 
thermally with the dust, and thereby greatly increase its 
cooling efficiency, by availing itself of the dust's 
continuum opacity. This circumstance defines a length-scale 
$\sim \, L\ci$ and a mass-scale 
$\sim \, M_\odot$ which are only weakly dependent on 
metallicity and environmental factors.

We concur with 
Larson's inference that dynamical fragmentation will then 
tend to deliver protostars with 
masses around the peak in the IMF at $0.4 M_\odot$. 

Dynamical fragmentation can only occur if the gas cools 
efficiently by radiating. Specifically, the gas must be 
able to radiate away -- on a dynamical timescale or faster -- 
the internal energy generated by gravitational compression 
(Hoyle 1953, Rees 1976, Low \& Lynden-Bell 1976, Silk 1977). 
At high densities this requires continuum opacity, 
and for cold  ($T \, \loo \, 10^2$K, say) gas having finite 
metallicity this opacity is supplied by dust.

Recently, Larson's analysis of the grouping of pre-MS stars in 
Taurus has been confirmed by Simon (1997). Simon has also 
repeated this analysis for the $\rho$ Ophiuchus and Orion-Trapezium 
star-formation regions. He finds that the surface-densities 
of neighbours and companions obey similar power laws, 
$N \, \propto \, \theta^{-0.5 \pm 0.2}$ on large scales, and 
$N \, \propto \, \theta^{-2.0 \pm 0.1}$ on small scales. 
However, he finds that the switch occurs at different projected 
separations: $\sim$ 0.06 pc in Taurus, $\sim$ 0.025 pc 
in $\rho$ Ophiuchus, and $\sim$ 0.002 pc in Orion-Trapezium. 

Simon argues that this variation is difficult to attribute to 
systematic changes in the Jeans length between the different 
star-formation regions. We propose that some of this variation  
arises because the different star-formation regions 
have very different large-scale surface densities. Proceeding 
from Taurus, through $\rho$ Ophiuchus, to Orion-Trapezium, 
the large-scale surface-density of stars increases significantly. 
and so the switch must shift to smaller projected separations, 
even if the intrinsic distribution of binary separations is identical.

Kitsionas, Gladwin and Whitworth (1998) have also re-analyzed 
Taurus, with a larger sample of pre-MS stars. They have 
repeated this analysis for the Chamaeleon I star-formation region. 
Power-law fits to $N(\theta)$ give exponents similar to those 
obtained by Larson (1995) and Simon (1997). For Chamaeleon I 
the switch is at $\sim$ 0.03 pc, as would be expected on the 
basis of its high overall surface-density. 

In Section 2 and Appendix A  we derive an expression for 
the compressional heating rate, ${\cal H}\co$, in a 
recently formed fragment. In Section 3 we evaluate thermal 
coupling between gas and dust. 
In Section 4 and Appendix B we derive an expression 
for the cooling rate, ${\cal L}\du$, due to thermal 
emission from dust. Appendix C demonstrates that for 
fragments having radius $R \la $ 0.1 pc, gas-cooling by 
dust dominates over gas-cooling by lines. In Section 5 we derive an 
expression for the mean (frequency-averaged) optical 
depth due to dust. In Section 6 we combine 
the results from the previous sections 
to define the domains on the ($R,M$) and ($\rho,T$) 
planes where gas-cooling by dust facilitates dynamical 
fragmentation. In Section 7 we 
discuss the results, and in Section 8 we summarize our 
main conclusions.

\section{Compressional heating}

For an approximately spherical fragment which has just 
become Jeans unstable, the mass $M\Je$, radius $R\Je$, 
temperature $T$ and mean gas-particle mass $\bar{m}$ 
are related by

\begin{equation}
\label{Jeans}
\frac{G \, M\Je}{R\Je} \; = \; 
\frac{{\cal J} k T}{\bar{m}} \; 
\equiv \; {\cal J} \, a\bg^2 \; ,
\end{equation}

\noindent where ${\cal J}$ is a numerical factor of order 
unity (see below), $k$ is Boltzmann's constant, and 
$a\bg \, = \, \left[ k T / \bar{m} \right]^{1/2}$ 
is the isothermal sound speed. 

If the gas remains isothermal, then by the time this 
fragment has contracted sufficiently for sub-fragments 
of mass ${\cal S} M\Je$ to become Jeans unstable, the 
original fragment has radius 
$R \; = \; {\cal S}^{2/3} R\Je$, 
and is collapsing at speed
$dR/dt \; = \; - \, {\cal M}({\cal S}) a\bg$. 
Consequently the compressional heating rate is

\[
{\cal H}\co \; = \; - \, P \; \frac{dV}{dt} \; = \; 
3 M\Je a\bg^2 \, \frac{\mbox{dln}[R]}{dt}
\]
 
\begin{equation}
\label{Hcomp}
\hspace{3.3cm} = \; \left\{ \frac{{\cal J} \, {\cal M}({\cal S})}
{{\cal S}^{2/3}} \right\} \; \frac{3 \, a\bg^5}{G} \; .
\end{equation}

\noindent In Appendix A we develop a simple time-dependent 
model of Jeans instability which enables us to estimate 
${\cal J} \; (\simeq \, 2.36)$, and 
${\cal M}({\cal S})$. For the purpose of numerical 
estimates, we set ${\cal S} \; = \; 0.2$ (up to 5 
fragments). 

For completeness, we also include a cosmic-ray heating rate 

\begin{equation}
\label{HCR}
{\cal H}_{\mbox{\tiny CR}} \; = \; \Gamma\bg \, M \; ,
\end{equation}

\noindent with $\Gamma\bg \, \simeq \, 2 \times 10^{-4}$ 
erg s$^{-1}$ g$^{-1}$. This corresponds to a cosmic-ray 
ionization rate ${\cal I}_{\mbox{\tiny CR}} \, \simeq \, 
6 \times 10^{-17}$ s$^{-1}$. Eqn. (\ref{HCR}) makes no allowance 
for the attenuation of cosmic rays in the interior of 
a dense fragment. However, this can only reduce the 
cosmic-ray heating rate, and it turns out that even the 
unattenuated rate is not very 
important in the regime with which we are concerned, 
i.e. the regime where gas-cooling by dust 
facilitates dynamical fragmentation.

\section{Thermal coupling between gas and dust}

If the gas in a fragment is thermally coupled to the dust, 
the continuum opacity of the dust affords the fragment an 
efficient means of radiating away the internal energy 
generated by gravitational compression. Here we formulate 
the rate ${\cal C}$ at which the gas transfers thermal 
energy to the dust. 

If a fraction $Z\du$ by mass of the interstellar medium 
is in the form of spherical dust grains having radius 
$r\du$ and internal density $\rho\du$, the rate of 
transfer of thermal energy to the dust is given by

\[
{\cal C}\gd \; = \; \frac{M}{\bar{m}} \, 
\left( \frac{8 k T}{\pi \bar{m}} \right)^{1/2} \, 
\frac{3 \rho Z\du}{4 r\du \rho\du} \hspace{1.2cm}
\] 

\begin{equation}
\label{Cdust}
\hspace{2cm} \times \, \left\{ 1 - \mbox{exp} \left[\frac{-75\mbox{K}}{T}\right] \right\} \, 
\frac{k (T-T\du)}{(\gamma - 1)} \; .
\end{equation}

\noindent For the purpose of numerical estimates we put 
$r\du \, = \, 10^{-5}$ cm and $\rho\du \, = \, 3$ g cm$^{-3}$. 
The term in braces, $\{\}$, represents the 
accommodation coefficient, i.e. the mean inelasticity 
of collisions between gas particles and dust grains (Hollenbach \& McKee 
1979). For the low temperatures ($T \, \loo \, 100$ K) 
with which we are mainly concerned here, 
$\gamma \, \simeq \, 5/3$, since the rotational degrees 
of freedom of H$_2$ are not excited.

\section{Radiative cooling}

The maximum radiative cooling rate is obtained by assuming 
that the fragment radiates like a blackbody, i.e.

\begin{equation}
{\cal L}_{\mbox{\tiny BB}} \; = \; 
4 \, \pi \, R^2 \, \sigma_{\mbox{\tiny SB}} \, T^4 \; .
\end{equation}

\noindent This is a maximum cooling rate on four counts.

First, it assumes that the collapsing fragment exists 
in isolation. In reality, part of the luminosity from 
the fragment will be cancelled by the flux of background 
radiation incident on the surface of the fragment. The 
background radiation field will consist of a cosmic 
component at $T\bg \, \sim$ 3 K, 
plus a contribution from local 
sources such as stars, protostars, circumstellar cocoons 
and discs, warm molecular clouds. A detailed evaluation 
of this contribution would be extremely model-dependent. For 
the purpose of this paper, we represent the background with 
a single blackbody component at temperature $T\bg \, \goo \,$ 3 K.

Second, it assumes that the cloud has continuum opacity 
at all wavelengths -- or, at the very least, across a 
broad range of wavelengths near the peak of the blackbody 
spectrum.

Third, it assumes that, at these wavelengths 
around the peak, scattering 
is negligible compared with absorption and emission. 

Fourth, it assumes that, at these wavelengths 
around the peak, the cloud is neither optically 
thin, nor very optically thick.

If we define a mean optical depth 
$\bar{\tau}(T)$ 
between the centre and the surface of the fragment 
(i.e. an optical depth averaged over the frequencies 
which contribute to radiative cooling of the fragment 
at temperature $T$), the radiative cooling rate is 
given approximately by

\[
{\cal L}\ra \; \simeq \; 
4 \, \pi \, R^2 \, \sigma_{\mbox{\tiny SB}} \; 
\left\{ \frac{T^4 \; \bar{\tau}(T) \tau\ci}
{(\bar{\tau}(T) + 1)(\bar{\tau}(T) + \tau\ci)} \right.
\]

\begin{equation}
\label{Lrad} 
\hspace{3cm} \left. \; - \; \frac{T\bg^4 \; \bar{\tau}(T\bg) \tau\ci}
{(\bar{\tau}(T\bg)+1) (\bar{\tau}(T\bg)+\tau\ci)} \right\} \; .
\end{equation}

\noindent Here, $\tau\ci$ is 
a critical optical depth, typically $\sim$ 20, and 
the second term in braces simply ensures that the cloud 
must be hotter than the background to cool radiatively. Eqn. 
(\ref{Lrad}) is constructed so as to give the correct 
asymptotic forms. For example, if  $T\bg \, = \, 0$, 
then in the optically thin limit, 

\[
{\cal L}\ra \; \longrightarrow \; 
4 \pi R^2 \sigma_{\mbox{\tiny SB}} T^4 \; \; \bar{\tau}(T) 
\; \; \; \; \; \; (\bar{\tau}(T) < 1) \; ;
\]

\noindent at modest optical depths, 

\[
{\cal L}\ra \; \longrightarrow \; 
4 \pi R^2 \sigma_{\mbox{\tiny SB}} T^4 \; \; \; \; \; \; 
(1 < \bar{\tau}(T) < \tau\ci) \; ;
\]

\noindent and at high optical depths,

\[ 
{\cal L}\ra \; \longrightarrow \; 
4 \pi R^2 \sigma_{\mbox{\tiny SB}} \; \; 
\frac{\tau\ci}{\bar{\tau}(T)} \; \; \; \; \; \; 
(\bar{\tau}(T) > \tau\ci) \; .
\]

\noindent This last form is derived in Appendix B, 
and takes account of the fact that high 
luminosities can only be transported through 
the interior of an optically thick fragment 
if there is a sufficiently large temperature 
difference between the centre of the fragment 
and its surface. To take account of this, we 
require that no more than a fraction $f$ of the 
fragment's mass be hotter than $FT$, where 
$T$ is now specifically the surface temperature. Hence 
the precise value of $\tau\ci$ depends on our 
choice of $f$ and $F$, and also on $\beta$ (the 
emissivity index): 
$\tau\ci \, = \, \tau\ci(f,F,\beta)$. 
Substituting $f \, = \, 1/3$, $F \, = \, 2$ 
(i.e. at least 2/3 of the fragment is below $2T$), 
we obtain $\tau\ci \, = \, 47.9 \; \& \; 30.8$ 
for $\beta \, = \, 1 \; \& \; 2$ respectively (see 
Appendix B). For the purpose of numerical estimates 
we shall adopt $\beta = 1.5$ and $\bar{\tau}\ci = 38.3$.

\section{Dust opacity}

The mean dust optical depth between the centre and the surface 
of a spherical, uniform-density fragment is

\[
\bar{\tau}(T\du) \; = \; \frac{9 M Z\du \bar{Q}(T\du)}
{16 \pi R^2 r\du \rho\du} \; ,
\]

\noindent where $\bar{Q}(T\du)$ is the Planck-mean absorption 
efficiency of a dust grain. Since our main concern here is 
with long wavelength radiation from dust at low temperatures, 
we adopt the prescription

\[
\bar{Q}(T\du) \; = \; 0.0004 \, 
\left[ \frac{T\du}{\mbox{10K}} \right]^\beta \; , \; \; \; \; \; 
1 \, \loo \, \beta \, \loo \, 2 \; .
\]

\noindent This corresponds to a mean mass-opacity coefficient
$\bar{\kappa}(T\du) \, = \, 3 Z\du \bar{Q}(T\du) / 
4 r\du \rho\du \, \simeq \, 10 \mbox{ cm}^2 
\mbox{ g}^{-1} \, Z\du \, \left[ T\du / \mbox{10K} \right]^\beta$, 
and hence a mean optical depth

\begin{equation}
\label{taubar}
\bar{\tau}(T\du) \; \simeq \; 2.4 \mbox{ cm}^2 \mbox{ g}^{-1} \; Z\du \;  
\frac{M}{R^2} \; \left[ \frac{T\du}{\mbox{10K}} \right]^\beta
\end{equation}

\noindent (cf. Hildebrand (1983), who recommends  
$Q(\lambda = 250 \mu \mbox{m}) \, = \, 0.0004$ and 
$\kappa(\lambda = 250 \mu \mbox{m}) \, = \, 0.1$ cm$^2$ g$^{-1}$). 

To account for dust evaporation, we multiply Eqn. 
(\ref{taubar}) by a factor 

\[
1 \; - \; 0.4 \, \mbox{exp} 
\left[ - \, \frac{100 \mbox{K}}{T\du} \right] 
\; - \; 0.6 \, \mbox{exp} 
\left[ - \, \frac{1000 \mbox{K}}{T\du} \right]  \; .
\]

\noindent The second term represents evaporation 
of the volatile grain mantle at temperatures of order 
100 K, and the third term represents evaporation 
of the refractory core at temperatures of order 1000 K. 
This rather {\it ad hoc} prescription is adequate for 
our purposes, since we are mainly concerned with gas-cooling 
by dust in the regime where the dust is cool and evaporation is negligible.

\section{Gas-cooling by dust during dynamical fragmentation}

The domain in which gas-cooling by dust facilitates 
quasi-freefall collapse and dynamical 
fragmentation (hereafter the {\it dust-domain}) is given 
by\footnote{The two heating processes 
operate simultaneously, and so their rates add in parallel. 
The two cooling processes operate sequentially, and so their 
rates add in series.}

\begin{equation}
\label{domain}
{\cal H}\co \, + \, {\cal H}\CR \; \loo \; 
\left\{ {\cal C}\gd^{-1} \, + \, {\cal L}\ra^{-1} \right\}^{-1} \; .
\end{equation}

\noindent The dust-domain can be defined on both the ($R,M$) and the 
($\rho,T$) planes, using Eqn. (\ref{Jeans}) and $\rho \, = \, 
3M / 4 \pi R^3$ to convert from one to the other. 

In order to simplify the analysis, we put $T\du \, = \, T\bg \, = \,$ 10 K in 
${\cal C}\gd$ (Eqn. (\ref{Cdust})), and $T\du \, = \, T$ in 
${\cal L}\ra$ (Eqn. (\ref{Lrad})). To justify this we note that  
the section of the dust-domain boundary where ${\cal C}\gd$ is important 
is quite distinct from the section where ${\cal L}\ra$ is important, 
{\it viz.}

Consider first the large-$R$ limit of the dust-domain, where 
$\, {\cal C}\gd \, \ll \, {\cal L}\ra \,$, so $\, {\cal C}\gd \,$ 
dominates the righthand side of Inequality (\ref{domain}), 
i.e. the critical process is the transfer of thermal energy 
from the gas to the dust. Under 
this circumstance the thermal balance of the dust involves a 
much more rapid energy turnover than that of the gas (e.g. 
Whitworth \& Clarke 1997). Hence 
the heat input from thermal coupling to the gas makes a 
negligible contribution to the thermal balance of the dust. 
The dust temperature $T\du$ is determined by the background 
radiation field, and adopts a value of order the background 
temperature $T\bg$. 

Next consider the small-$R$ limit of the dust-domain, where 
$\, {\cal C}\gd \, \gg \, {\cal L}\ra \,$, so $\, {\cal L}\ra \,$ 
dominates the righthand side of Inequality (\ref{domain}), i.e. 
the critical process is the emission of radiation by the dust and 
its escape from the fragment. At these high densities the gas and dust are closely coupled. A very small temperature difference suffices to transfer the 
compressional heating of the gas to the dust, and we can approximate 
$T\du \, \simeq \, T$. 

In between these two limits, the dust-domain boundary is determined at 
high temperatures by dust evaporation, and at low temperatures by 
the fact that the gas cannot cool below the background temperature. 
Consequently there is no need to formulate accurately the 
circumstances under which ${\cal C}\gd \, \sim {\cal L}\ra$. 

\begin{figure}
\plotone{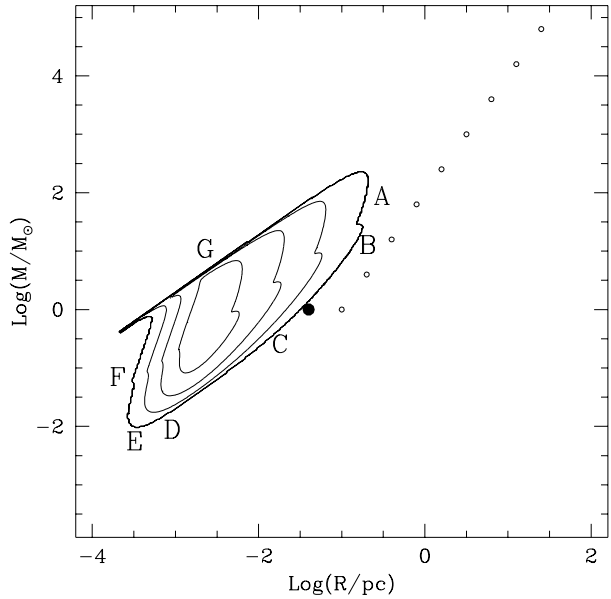}
\caption{The ($R,M$) plane. The outer contour is the limit of 
the dust-domain, where gas-cooling by dust enables the gas to fragment 
dynamically. The other contours show where the ratio of 
compressional heating to gas-cooling by dust equals 0.32, 0.10 \& 0.03. 
The letters refer to the different processes limiting the 
dust-domain, as discussed in the text. The line of open circles represents 
the average observed properties of clumps. The large dot marks the 
point (0.04 pc,$M_\odot$), i.e. the critical entity identified by 
Larson.}
\end{figure}

Figure 1 shows contours of constant 

\[
{\cal R} \; \equiv \; 
({\cal H}\co + {\cal H}\CR) \, 
({\cal C}\gd^{-1} + {\cal L}\ra^{-1}) 
\]

\noindent on the ($R,M$) plane, for $\beta \, = \, 1.5$, 
$T\bg \, =$ 10K, and $Z\du \, = \, 0.01$. The region inside the 
outer contour is the dust-domain. Contours are drawn for 
${\cal R} \, =$ 1.00, 0.32, 0.10 \& 0.03, to emphasize how efficient 
gas-cooling by dust becomes inside the dust-domain. 

Starting at the righthand side, 
and working round in a clockwise sense, the dust-domain limits are 
caused by (A) gas-dust coupling in the high-temperature (elastic) limit; 
(B) gas-dust coupling in the low-temperature (perfectly inelastic) 
limit; (C) the background temperature acting as a minimum temperature 
($T \, \geq \, T\bg$); (D) the dust being optically thin; (E) 
the dust optical depth being intermediate 
($1 \loo \bar{\tau} \loo \tau\ci$); (F) the dust being optically thick; 
(G) dust evaporation. {\it Partial} dust evaporation actually promotes 
fragmentation in the high density limit by reducing the dust 
optical depth, but it inhibits gas-dust coupling; this is why the 
dust-domain boundaries curve towards higher densities as they approach the evaporation limit. The same contours are presented on the ($\rho,T$) 
plane in Figure 2. The dashed line indicates the background temperature 
$T\bg$.

\begin{figure}
\plotone{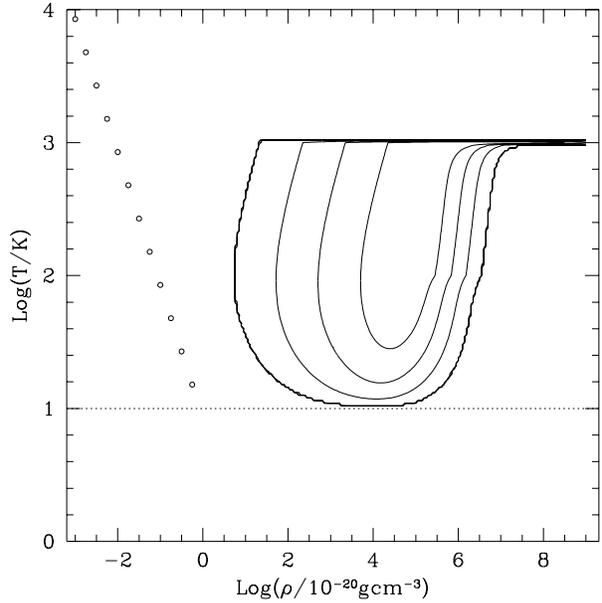}
\caption{The ($\rho,T$) plane. The outer contour is the limit of 
the dust-domain, where gas-cooling by dust enables the gas to fragment 
dynamically. The other contours show where the ratio of 
compressional heating to gas-cooling by dust equals 0.32, 0.10 \& 0.03. 
The line of open circles represents the average observed properties of 
clumps. The dashed line represents the background temperature.}
\end{figure}

The most relevant part of the dust-domain is probably the part corresponding to low temperatures, $T \la $50K, but we have plotted the whole domain in order to avoid the extremely model-dependent complexities of predicting the gas temperature precisely.

Of course, to be important in this context gas-cooling by dust must not only be able to handle the net heating of a fragment, it must also be able to handle it more effectively than, or at least as effectively as, gas-cooling by lines. Figure 1 shows that the dust-domain has $R \la $0.1pc, and Appendix C demonstrates that for protostellar fragments with $R \la $0.1pc, gas-cooling by dust is more effective than gas-cooling by lines.

\section{Discussion}

On Figure 1, we have also plotted the mean observed distribution of 
clump masses and radii, as collated by Chi\`eze (1987). This 
distribution approximates to

\[
M \; \simeq \; M_\odot \, \left[ \frac{R}{0.1 \mbox{pc}} \right]^2 \; .
\]

\noindent and is marked by a line of open circles on Figure 1. 
If the clumps were thermally virialized, we could use Eqn. 
(\ref{Jeans}) to convert this distribution into

\[
T \; \simeq \; 8.5 \, \mbox{K} \left[ \frac{\rho}
{10^{-20} \mbox{ g cm}^{-3}} \right]^{-1} \; ;
\]

\noindent and this is marked by the line of open circles on Figures 2.

However, all but the very low-mass clumps are 
supported by non-thermal motions, with the gas being at rather 
low temperatures, typically in the range 10 to 15 K, and seldom 
above 50K. Thus, although the locus of clumps in Figure 1 appears to run 
rather close to the boundary of the dust-domain, it is only the 
low-mass clumps -- those supported mainly by thermal pressure -- 
that are really close to the dust-domain. Even these may be unable 
to access the dust-domain without some external perturbation. 

The gas in a high-mass clump can probably only enter the dust-domain 
if it is compressed and heated by 
a shock, for example following a collision with another clump of 
comparable mass, or as a consequence of being overrun by an expanding 
nebula (HII region, stellar-wind bubble or supernova remnant). Gas-cooling by lines will play an important role in such shocks (as a means of removing the dissipated energy), but Appendix C shows that the shock-compressed gas will then only be able to condense further and fragment if it is dense enough to couple thermally to the dust.

A number of observations might be explained on the basis of the 
dust-domain and its relation to the locus of observed clumps.

\subsection{Low-mass clumps}

Low-mass ($\sim M_\odot$) clumps are supported mainly by thermal 
pressure (Myers 1983). Therefore, in the absence of external perturbations, they 
should relax to hydrostatic equilibrium. However, Figures 1 \& 2 
suggest that such clumps need a relatively small external 
perturbation -- and possibly no perturbation at all -- 
to enter the dust-domain. Consequently we might expect a large fraction of stars to be formed from the collapse of such clumps. If dynamical 
fragmentation leads typically to the formation of a binary or 
triple system, with most of the clump mass ending up in the 
component stars, then the Initial Mass Function (IMF) should peak 
in the range 0.2 to 0.5 $M_\odot$, as it apparently does (Scalo 1986). 
This last point was emphasized by Larson (1995).

Moreover, if an external perturbation is involved, this should 
greatly increase the chances of permanent fragmentation during 
the ensuing collapse. This is because the clump would be delivered 
into the dust-domain with highly non-linear substructure, and gravity 
would then rapidly amplify this non-linear substructure during the subsequent quasi-freefall collapse. 

By contrast, if it is possible for a clump to condense quasistatically 
into the dust-domain, it arrives there in a rather too well organized 
state, and collapse seems likely to lead to a single star. Under this 
circumstance, binary formation may require some sort of 
intrinsic instability, for example in a circumstellar disc.

Support for the notion that even low-mass clumps either require an 
external perturbation to trigger their collapse, or condense 
rather quasistatically into the dust-domain before collapsing to form 
stars, comes from the observation (Myers, private communication) 
that the majority of starless low-mass clumps are located in the outer 
reaches of Taurus, where presumably external perturbations are few 
and far between. 

Further support comes from the observation that in low-mass clumps 
associated with IRAS sources, the IRAS source (i.e. the putative 
newly-formed star) is normally significantly displaced from the 
centre of the (presumably placental) clump (Beichman et al. 1986, 
Clark 1987).

\subsection{High-mass clumps}

High-mass ($\goo 10 M_\odot$) clumps are supported mainly by non-thermal 
motions. From Figure 2, it appears that they are unable to 
enter the dust-domain, 
without a substantial external impulse -- most likely a shock, due 
to a collision with another comparable clump, or due to being overrun 
by an expanding nebula. 
A sufficiently strong shock would compress the gas to the point 
where it became thermally coupled to the dust. If our hypothesis is correct, this would lead to efficient fragmentation and the formation of a star-cluster.

There is evidence that molecular clouds have an hierarchical structure, 
i.e. clumps nested within clumps. If this is so, a high-mass clump 
should be envisaged as an ensemble of lower-mass clumps. Consequently, 
even when star formation is triggered in a high-mass clump (by an external 
perturbation), most of the gas ending up in stars will probably 
reside in low-mass ($\sim M_\odot$) subclumps, and this should lead to a 
majority of the stars formed having masses in the range 0.2 to 0.5 $M_\odot$.

\subsection{$M\mi$ and $M\ma$} 

Several authors (Rees 1976, Low \& Lynden-Bell 1976, Silk 1977) 
have determined the minimum mass for star 
formation $M\mi$ on the basis of opacity-limited fragmentation. 
With reference to Figure 1, this is equivalent to identifying $M\mi$ 
with the smallest mass falling inside the dust-domain, and it yields a 
value $M\mi \, \sim \, 0.01 M_\odot$. However, to reach this part of 
the dust-domain from the conditions obtaining inside a clump would 
require either (i) several consecutive fragmentation episodes, or (ii) 
a single fragmentation episode generating a very large number of pieces. 
Numerical simulations (Larson 1978) suggest that dynamical fragmentation 
is a one-stage process, i.e. that during freefall gravity only amplifies 
the substructure which is present at the outset. This implies that 
(i) cannot work, and that (ii) requires an external perturbation 
with improbably high power on high spatial frequencies. Therefore we 
do not expect many stars with masses $\sim$ $0.01 M_\odot$.

The largest mass falling inside the dust-domain is 
$M\ma \, \sim \, 100 M_\odot$. 
We speculate that this is because clumps having higher 
mass cannot (coherently) enter the dust-domain, no matter what external perturbation they experience, and therefore cannot collapse to form 
stars. 

\begin{figure}
\plotone{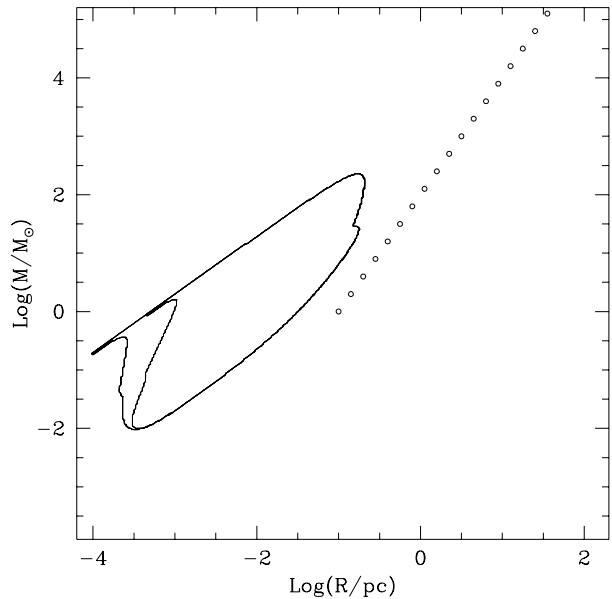}
\caption{The ($R,M$) plane. The effect on the dust-domain of 
varying the far-infrared emissivity index $\beta$. The heavy 
contour delimits the dust-domain when $\beta = 1$. The light 
contour delimits the dust-domain when $\beta = 2$. Note that 
the two contours only diverge at small $R$.}  
\end{figure}

\subsection{Dependence on $\beta$, $T\bg$ and $Z\du$}

In Figure 3 we show how the dust-domain is altered on the ($R,M$) 
plane, if one adopts 
a different far-infrared emissivity index for the dust, i.e. 
$\beta \, = \, 1$ or $\beta \, = \, 2$. The only significant change 
is on the small-$R$ boundary of the dust-domain, where the dust is 
optically thick to its own radiation. Since we are here more 
concerned with the large-$R$ boundary, this variation will not 
affect our conclusions significantly.

\begin{figure}
\plotone{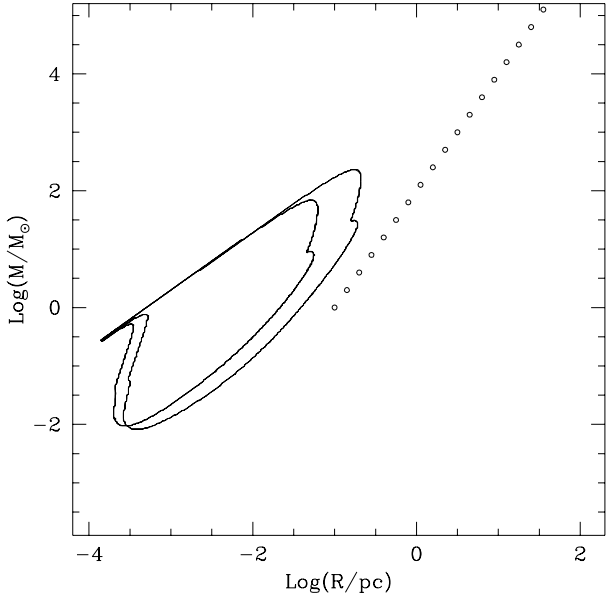}
\caption{The ($R,M$) plane. The effect on the dust-domain of 
reducing the metallicity to $Z\du = 0.003$, heavy contour; 
and reducing the background temperature to $T\bg = 5$ K, light 
contour.} 
\end{figure}

In Figure 4 we show how the dust-domain is altered on the ($R,M$) 
plane, if one adopts a different background temperature, i.e. 
$T\bg \, = \,$ 5 K (instead of $T\bg \, = \,$ 10 K), or a different 
metallicity, i.e. $Z\du \, = \,$ 0.003 (instead of $Z\du \, = \,$ 0.01. 
Changing the background temperature has little effect on the 
boundary of the dust-domain where it is close to the locus of clumps.

Reducing the metallicity moves the dust-domain to smaller 
$R$-values. Unless the locus of clumps is also different, 
in the sense of being characterized by a larger mean column-density, 
this increases the gap between the dust-domain and 
the locus of clumps. Consequently, where 
the metallicity is well below solar, it should be much harder for 
star formation to occur spontaneously. This might mean that 
star formation is more dependent on an external impulse, 
and hence that star formation is more likely to occur in bursts. 
This could dictate the pattern of star formation in a galaxy having low metallicity, for example a young galaxy or an irregular galaxy -- in the sense of making star formation more burst-like.

\section{Conclusions}

We have evaluated the conditions under which gas-cooling by dust 
enables a clump to collapse and fragment dynamically. We describe these conditions as the {\it dust-domain}. We have also shown that in the dust-domain gas-cooling by dust dominates over gas-cooling by dust. 

The observed properties of clumps 
in star-forming molecular clouds are such that low-mass ($\sim M_\odot$) 
clumps are close to the dust-domain. Consequently  
only a small external perturbation is needed to propel them into 
the dust-domain -- and possibly no perturbation at all. Therefore, 
if such clumps fragment into binary or triple systems, with most of their 
mass ending up in the components of these systems, they will spawn stars  
having masses concentrated in the range 0.2 to 0.5 $M_\odot$, which 
is the peak of the IMF (cf. Larson 1995).

In isolated quiescent regions, low-mass cores may survive for 
long periods, for lack of a sufficient external impulse to trigger 
(or accelerate) their collapse. They would be observed as starless 
clumps. 

High-mass clumps require a substantial external impulse to reach the 
dust-domain and start collapsing. Collisions with other clumps and/or 
shocks due to expanding nebulae can provide a sufficient impulse. 
This may be why star formation is (apparently) more efficient 
in dynamically agitated regions.

The dust-domain may also define the 
minimum and maximum masses for star formation, 
$M\mi \, \sim \, 0.01 M_\odot$ and $M\ma \, \sim \, 100 M_\odot$, 
but the majority of stars should fall well within these limits.

In regions of low metallicity, for instance irregular galaxies, 
star formation may be more prone to occur in bursts.

\vspace{0.8cm}

{\noindent\Large\bf Appendix A. Isothermal fragmentation}

\vspace{0.5cm}

\noindent Consider a uniform medium with density $\rho\bg$ 
and isothermal sound speed $a\bg$. Now suppose 
that a spherical fragment having radius $R\bg$, and hence 
mass $M\bg \, = \, 4 \pi R\bg^3 \rho\bg / 3$, 
attempts to condense out of the background by 
contracting homologously, so that it remains 
spherical with instantaneous radius $R$, and 
uniform with instantaneous density $\rho \, = \, \rho\bg \, 
(R/R\bg)^3$. Further suppose that the isothermal 
sound speed is constant, $a \, = \, a\bg$, and 
that the external medium exerts a constant pressure 
$P_{\mbox{\tiny ext}} \, = \, \rho\bg \, a\bg^2$ at 
the boundary of the fragment. 
Radial excursions of the fragment are controlled by 
a potential function $\phi(R)$, in the sense that 
$d^2R/dt^2 \; = \; - \, d\phi/dR$ and 
$(dR/dt) \; = \; \left[ -2\phi(R) \right]^{1/2}$. 
$\phi(R)$ is derived (Whitworth 1981) from the 
energy equation

\begin{equation}
\label{energy}
\frac{d}{dt} \left\{ {\cal K} + {\cal G} \right\} \; = \; 
\left[ P_{\mbox{\tiny int}} - P_{\mbox{\tiny ext}} \right] 
\frac{dV}{dt} \; ,
\end{equation}

\noindent where ${\cal K} \, = \, (3/10) M\bg (dR/dt)^2$ 
is the radial kinetic energy, ${\cal G} \, = \, - \, 
3 G M\bg^2 / 5 R$ is the self-gravitational potential 
energy, and $P_{\mbox{\tiny int}} \, = \, 3 M\bg a\bg^2 
/ 4 \pi R^3$ is the internal pressure. $\phi(R)$ can be 
written in the form

\[
\frac{\phi(R)}{a\bg^2} \; = \; 
- \, \frac{15}{2^{8/3}} \, \left( \frac{R\bg}{R\Je} \right)^2 \, 
\left[ \left( \frac{R}{R\bg} \right)^{-1} - 1 \right]
\]

\begin{equation}
\label{potfn} 
\hspace{2.5cm} - \, 5 \, \mbox{ln} \left( \frac{R}{R\bg} \right) \; + \, 
\frac{5}{3} \, \left[ \left( \frac{R}{R\bg} \right)^3 
- 1 \right] \; ,
\end{equation}

\noindent where the first term on the righthand side 
represents self-gravity, the second term internal 
pressure, the third term external pressure, and

\begin{equation}
\label{RJeans}
R\Je \; = \; \frac{3}{2^{7/3}} \, 
\left( \frac{5}{\pi G \rho\bg} \right)^{1/2} \, a\bg \; .
\end{equation}

\noindent Stable equilibrium states ($d\phi/dR \, = \, 0, 
\; d^2\phi/dR^2 \, > 0$) exist only for 
$R\bg \, < \, R\Je$; for $R\bg \geq R\Je$ the fragment 
contracts indefinitely. Therefore $R\Je$ can be identified as 
the Jeans radius. The Jeans mass is given by 

\begin{equation}
\label{MJeans}
M\Je \; = \; \frac{4 \pi R\Je^3 \rho\bg}{3} \; = \; 
\frac{3^2 \, \pi}{2^5} \, 
\left( \frac{5}{\pi \, G} \right)^{3/2} \, 
\frac{a\bg^3}{\rho\bg^{1/2}} \; ,
\end{equation}

\noindent and so

\begin{equation}
\label{JJeans}
{\cal J} \; \equiv \; \frac{G \, M\Je}{R\Je \, a\bg^2} \; 
= \; \frac{15}{2^{8/3}} \; \simeq \; 2.3623 \; .
\end{equation}

If we consider a marginally Jeans-unstable fragment with 
mass $M\Je$, starting to condense out from radius $R\Je$, 
then by the time sub-fragments with mass ${\cal S} M\Je$ 
become Jeans unstable, the radius of the initial fragment 
has decreased to ${\cal S}^{2/3} R\Je$. From Eqn. 
(\ref{potfn}), it follows that the Mach Number of the 
contraction is given by

\[
{\cal M}({\cal S}) \, \equiv \, - \, \frac{dR/dt}{a\bg} \,  = \, 
\left\{ \frac{15}{2^{5/3}} \, \left[ {\cal S}^{-2/3} - 1 \right] \right.
\]

\begin{equation}
\label{Mach} 
\hspace{3.4cm} \left. + \, \frac{20}{3} \mbox{ln} \left[ {\cal S} \right] 
\, + \, \frac{10}{3} \left[ 1 - {\cal S}^2 \right] \right\}^{\frac{1}{2}}.
\end{equation}

\vspace{0.1cm}

Representative values of ${\cal M}({\cal S})$ 
are tabulated below

\begin{center}
\begin{tabular}{ccccc}
${\cal S}$ & & ${\cal M}$ & & ${\cal M}/{\cal S}^{2/3}$ \\
0.5 & & 0.81 & & 1.41 \\
0.4 & & 0.82 & & 1.51 \\
0.3 & & 0.91 & & 2.03 \\
0.2 & & 1.25 & & 3.65\\
0.1 & & 2.27 & & 10.54 \\ 
\end{tabular}
\end{center}

\noindent The slow variation of ${\cal M}$ with ${\cal S}$ 
for ${\cal S} \, \sim \, 0.5$ reflects the fact that 
the point of inflexion ($d^2\phi/dR^2 \, = \, 0$) 
which defines the marginally Jeans-unstable fragment 
occurs at ${\cal S} \, = \, 0.5$, $R \, = \, 
{\cal S}^{2/3} R\Je \, \simeq \, 0.63 R\Je$. Consequently 
acceleration of the collapse is slow at this stage.

There is apparently an inconsistency in the above analysis, in 
the sense that -- 
strictly speaking -- a marginally Jeans-unstable fragment would be 
centrally condensed, and hence the value of ${\cal J}$ would be 
smaller than the one we have derived. However, since here we are 
mainly concerned with 
dynamical fragmentation, we do not envisage there being sufficient 
time for the fragment of the moment to relax towards detailed 
hydrostatic balance. Therefore it is appropriate to use the 
results we have derived on the basis of a uniform-density fragment.

\vspace{0.8cm}

{\noindent\Large\bf Appendix B. Radiative transport 

\noindent inside an optically thick fragment}

\vspace{0.5cm}

\noindent In order to estimate the temperature difference 
required to drive radiative energy transport from the centre 
of the fragment to its surface in the high optical depth limit, 
we introduce a dummy optical depth variable $\tau'$ with 
$\tau' = 0$ at the centre and $\tau' = \bar{\tau}(T)$ 
at the surface. For this calculation, we must allow that the 
temperature is a function of position $\theta(\tau')$, and 
increases from $\theta = T$ at the 
surface to higher values in the interior. We shall adopt the 
condition that isothermality breaks down once a fraction $f$ 
of the fragment's mass is heated above $FT$, and adopt 
$f = 1/3$, $F = 2$.

Since the mean dust opacity depends on 
temperature (see Eq. (\ref{taubar})), 
the effective optical-depth differential is 

\[
(\theta/T)^{\beta} \; d \tau' \; ,
\]

\noindent and so the flux is given by

\[
F(\tau') \; \simeq \; 
- \; \frac{4}{3} \; \frac{d}{(\theta/T)^{\beta} d \tau'}
\left[ \sigma \theta^{4}(\tau') \right] \; . 
\]

For simplicity we assume that heat is deposited uniformly throughout the fragment, so the divergence of the flux must satisfy

\begin{eqnarray}
\nabla . {\bf F} & \equiv & \frac{1}{r^2} \, \frac{d}{dr} \left\{ r^2 F(r) \right\} \nonumber \\
 & \equiv & - \, \frac{16 \sigma T^\beta \bar{\tau}(T)}{3R} \, \frac{1}{\tau'^2} \, 
\frac{d}{d\tau'} \left\{ \tau'^2 \theta^{(3 - \beta)}(\tau') 
\frac{d\theta}{d\tau'} \right\} \nonumber \\
 & = & \frac{3{\cal H}\co}{4 \pi R^3} \nonumber \\
\end{eqnarray}

\noindent where we have substituted $r = R \tau' / \bar{\tau}(T)$ and 
$dr = R d\tau' / \bar{\tau}(T)$.

Integrating this equation, and applying the boundary conditions $d\theta/d\tau' = 0$ at $\tau' = 0$, and $\theta = T$ 
at $\tau' = \bar{\tau}(T)$, we obtain 

\begin{eqnarray}
\theta^{(4 - \beta)}(\tau') & = & T^{(4 - \beta)} \nonumber \\
 & + & \frac{3 (4 - \beta) {\cal H}_{\mbox{\tiny comp}} \bar{\tau}(T)}
{128 \pi R \sigma T^\beta}  
\left\{ 1 - \left[ \frac{\tau'}{\bar{\tau}(T)} \right]^2 \right\} \, . \nonumber \\
\end{eqnarray}

Substituting $\tau'/\bar{\tau}(T) = f^{1/3}$ and $\theta(\tau') 
< FT$, gives the condition for approximate isothermality to break down:

\begin{eqnarray}
\frac{{\cal H}_{\mbox{\tiny comp}} \bar{\tau}(T)}{4 \pi R^2 \sigma T^4} & < & \tau_{\mbox{\tiny crit}}(f,F,\beta) \nonumber \\
 & & \hspace{0.3cm} = \; \frac{32 \left[ F^{(4 - \beta)} - 1 \right] }
{3 \left[4 - \beta\right] \left[ 1 - f^{2/3} \right]} \; . \nonumber \\
\end{eqnarray}

\noindent For $f = 1/3$ and $F = 2$, we obtain the values

\begin{center}
\begin{tabular}{lclll}
\hspace{1cm} $\beta$ & \hspace{0.5cm} = \hspace{0.5cm} & 1.00 \hspace{0.5cm} & 1.50 \hspace{0.5cm} & 2.00 \\
\hspace{1cm} $\tau_{\mbox{\tiny crit}} \; \; \;$ & $\simeq$ & 47.9 & 38.3 & 30.8 \\
\end{tabular}
\end{center}

\vspace{0.8cm}

{\noindent\Large\bf Appendix C. \hspace{0.3cm} Gas-cooling by dust

\noindent versus gas-cooling by lines}

\vspace{0.5cm}

\noindent Consider a molecule with mass $m\X$ and moment of inertia $I\X$. 
Its rotational energy levels are at $E_J \, = \, J (J+1) 
\hbar^2 / 2 I\X \, \simeq \, J^2 \hbar^2 / 2 I\X$, and they 
will be significantly excited up to $E_{\Jh} \, = \, {\cal F}kT\;$
(where ${\cal F}$ is a factor of order unity), so we can put

\[
\Jh \; \simeq \; 
\frac{\left[ 2 {\cal F} I\X k T \right]^{1/2}}{\hbar} \; .
\]

The rotational lines have 

\[
h \nu_{_{(J \rightarrow J-1)}} \; = \; \frac{J \hbar^2}{I\X} 
\; \; \; \; \; \longrightarrow \; \; \; \; \; 
\nu_{_{(J \rightarrow J-1)}} \; = \; \frac{J \hbar}{2 \pi I\X}
\]

\noindent and hence 

\[
h \nu_{_{(J \rightarrow J-1)}} \; \leq \; 
h \nu_{_{(\Jh \rightarrow \Jh-1)}} \; \simeq \; 
\frac{2 {\cal F} k T}{\Jh} \; .
\]

To calculate the maximum possible cooling rate due to 
this molecule, we assume that all the lines up to and including 
$(\Jh \, \rightarrow \, \Jh - 1)$ are optically thick. 

Under this circumstance, the intensity in a line is given by 
the Planck Function. Again, since we are after a maximum 
possible cooling rate, we can replace the Planck Function with 
the Rayleigh-Jeans approximation, $B_\nu(T) \, \simeq \, 
2 k T \nu^2 / c^2$. Since the peak of the Planck Function is at 
$h \nu \, \simeq \, 5 k T$, whereas the highest frequency 
optically thick cooling line is at 
$h \nu_{_{(\Jh \rightarrow \Jh-1)}} \; \simeq \; 
2 {\cal F} k T/ \Jh$, 
this is actually a fairly good approximation as long as 
$\Jh \, > \, 0.4 {\cal F}$.

The width of a line is determined by the velocity dispersion. 
For the thermally supported low-mass clumps with which we are 
concerned the velocity width is given by

\[
\Delta v \; \sim \; 
4 \, \left[ \frac{2 k T}{\pi m\X} \right]^{1/2} \; = \; 
4a\bg \, \left[ \frac{2 \bar{m}}{\pi m\X} \right]^{1/2} \; .
\]

If we consider a clump of mass $M$ and radius $R$, then the 
maximum cooling rate due to a single line from this 
particular molecule is 

\begin{eqnarray}
{\cal C}_{_{J}} & \simeq & \left. 
4 \pi R^2 \; \pi \left[ B_\nu(T) - B_\nu(T\bg) \right] \; 
\frac{\nu \Delta v}{c} \right|_{\nu = J \hbar / 2 \pi I\X} \\
 & \simeq & \frac{4 R^2 k \left[ T - T\bg \right] a}{\pi} \, 
\left[ \frac{2 \bar{m}}{\pi m\X} \right]^{1/2} \, 
\left[ \frac{\hbar J}{c I\X} \right]^3 \; .
\end{eqnarray}

\noindent where $T\bg$ is the temperature of the background 
radiation field.

The total cooling rate due to the molecule is 
obtained by summing over all the lines:

\begin{eqnarray}
{\cal C}\X & \simeq & 
\sum_{J = 1}^{J = \Jh} \, \left\{ {\cal C}_{_{J}} \right\} \\
 & \simeq & \frac{R^2 k \left[ T - T\bg \right] a}{\pi} \, 
\left[ \frac{2 \bar{m}}{\pi m\X} \right]^{1/2} \, 
\left[ \frac{\hbar }{c I\X} \right]^3\, \frac{\Jh^4}{4} \\
 & \simeq & \frac{4 {\cal F}^2 R^2 \bar{m}^2 k[T-T\bg] a^5}
{\pi c^3 \hbar I\X} \, \left[ \frac{2 \bar{m}}
{\pi m\X} \right]^{1/2} \; .
\end{eqnarray}

From the results of Neufeld, Lepp \& Melnick (1995), we 
infer that at high densities ({\it i.e.} in the optically 
thick limit) $^{12}$C$^{16}$O contributes $\sim$ 4 
$\%$ of the net cooling, so we put $m\X \, = \, m\CO \, 
\simeq \, 5 \times 10^{-23}$ g, $I\X \, = \, I\CO \, 
\simeq \, 6 \times 10^{-38}$ g cm$^2$ s$^{-1}$, and write 
the net cooling rate in the form 

\begin{eqnarray}
{\cal C}_{\mbox{\tiny line}} & \simeq & 
{\cal N} {\cal C}\CO \nonumber \\
 & \simeq & 
\frac{4 {\cal N} {\cal F}^2 R^2 \bar{m}^2 k[T-T\bg] a^5}
{\pi c^3 \hbar I\CO} \, \left[ \frac{2 \bar{m}}
{\pi m\CO} \right]^{1/2} \; . \nonumber \\
\end{eqnarray}

\noindent where ${\cal N} \, \simeq \, 25$ is intended to represent the contribution from 24 other similar molecular lines from different molecules and/or isotopic variants. This net gas-cooling rate due to (molecular) lines should be compared with the gas-cooling rate due to dust.

It turns out that the two cooling rates are comparable 
in the relatively low density regime where the  
gas-cooling rate due to dust is dominated by the rate of transfer of 
thermal energy from the gas to the dust, i.e.

\begin{eqnarray}
{\cal C}_{\mbox{\tiny g-d}} & = & 
\frac{M}{\bar{m}} \left[ \frac{8 k T}{\pi \bar{m}} \right]^{1/2} 
\frac{3 \rho Z\du}{4 r\du \rho\du} \nonumber \\
 & & \times \; \left\{ 1 - \mbox{exp} \left[ \frac{-75 \mbox{K}}{T} \right]\right\} \frac{k [T - T\du]}{[\gamma - 1]} \, , \nonumber \\
\end{eqnarray}

\noindent rather than the rate at which the dust reradiates this energy 
(${\cal L}\ra$). Additionally, at the low temperatures with which we 
are mainly concerned, we can simplify the analysis by 
setting the accommodation coefficient (the term in braces, 
$\{ \; \}$) to unity.

Putting 

\[
\frac{k T}{\bar{m}} \; = \; a^2 \; = \; \frac{G M}{{\cal J} R} \; ,
\]

\noindent the condition for line cooling to dominate over 
dust cooling becomes

\begin{eqnarray}
R > R_{\mbox{\tiny crit}} & \simeq &
\left[ \frac{9 {\cal J}^2 \hbar Z\du I\CO 
\left[ T - T\du \right] }{32 {\cal N} {\cal F}^2 
G^2 r\du \rho\du \left[ \gamma - 1 \right] 
\left[ T - T\bg \right]} \right]^{1/3} \nonumber \\
 & & \times \; \left[ \frac{m\CO}{\bar{m}} \right]^{1/6} 
\frac{c}{\bar{m}} \, . \nonumber \\
\end{eqnarray}

\noindent With ${\cal J} \, = \, 2.4$, $Z\du \, = \, 0.01$, 
${\cal N} \, = \, 25$, ${\cal F} \, = \, 2$, $r\du \, = \, 
10^{-5}$ cm, $\rho\du \, = \, 3$ g cm$^{-3}$, $\gamma \, 
= \, 5/3$, and $\bar{m} \, = \, 4 \times 10^{-24}$ g, 
this reduces to

\begin{equation}
R \; > \; R_{\mbox{\tiny crit}} \; \simeq \; 0.3 \, \mbox{pc} \; 
\left[ \frac{ T - T\du }
{ T - T\bg } \right]^{1/3} \; .
\end{equation}

With a view to making $R_{\mbox{\tiny crit}}$ as small as possible, we put $(T-T\bg) \simeq 10$K (e.g. $T\bg \, \simeq \, 3$K and $T \, \simeq \, 13$K). Then to reduce $R_{\mbox{\tiny crit}}$ to 0.1 pc requires $(T-T\du) < 0.3$K, and to reduce it to 0.03 pc requires $(T-T\du) < 0.01$K. It would be an extraordinary coincidence if, {\it in a regime where gas-cooling by lines was dominant}, thermal balance were to deliver such a small value of $(T-T\du)$. We conclude that for fragments with $R \la$ 0.1 pc, gas-cooling is dominated by dust.

\vspace{0.8cm}

{\noindent\large\bf ACKNOWLEDGEMENTS}

\vspace{0.4cm}

\noindent HMJB is supported by PPARC grant GR/K94157, and NF is 
supported by a PPARC studentship. We thank Cathie Clarke and Phil 
Myers for useful discussions.

\end{document}